\newcommand{\bolding}{1}
\documentclass[authoryear,12pt]{article}
\usepackage[a4paper, total={6.5in, 10in}]{geometry}
\usepackage{amsmath,amssymb,amsfonts,amsthm,array,nccmath,pxfonts}%
\usepackage{booktabs}
\usepackage{graphicx}%
\usepackage{multirow}%
\usepackage{bm}%
\usepackage{mathrsfs}%
\usepackage[title]{appendix}%
\usepackage{xcolor}%
\usepackage{textcomp}%
\usepackage{manyfoot}
\usepackage{microtype}%
\usepackage{booktabs}%
\usepackage{algorithm}%
\usepackage{algorithmicx}%
\usepackage{algpseudocode}%
\usepackage{listings}%
\usepackage{pxfonts}%
\usepackage{xcolor}%
\usepackage{subcaption}
\usepackage{natbib}
\usepackage[font = small]{caption}
\usepackage[colorlinks=true,linkcolor=black, citecolor=blue, urlcolor=blue]{hyperref}
\usepackage{cleveref}%

            \title{Bayesian Wombling for Tracking Invasive Species}
            \author{Aritra Halder$^a$, Joseph R. Stinziano$^b$, Joshua Persi$^b$
            		\\
            			$^a$Department of Biostatistics \& Epidemiology,\\
            			Drexel University,
            			Philadelphia, PA, USA.\\
                            $^b$Plant Data, Analytics, and Modeling Unit\\
                            Canadian Food Inspection Agency, Government of Canada.
                            }
            \date{}

\begin{document}
\newcommand{\T}{\mathrm{\scriptscriptstyle T} }
\newcommand{\trans}{^\top} 
\newcommand{\prm}{^{\prime}} 
\newcommand{\dprm}{^{\prime\prime}} 
\newcommand{\fim}[1]{\trans{\bX}#1\bX} 
\newcommand{\dd}[1]{\mfrac {\mathrm{d}}{\mathrm{d}#1}} 
\newcommand{\del}[2]{\frac{\partial#1}{\partial #2}} 
\newcommand{\deltwo}[1]{\mfrac{\partial^2}{\partial #1^2}} 
\newcommand\scalemath[2]{\scalebox{#1}{\mbox{\ensuremath{\displaystyle #2}}}}
\newcommand{\beginsupplement}{%
        \setcounter{equation}{0}
        \renewcommand{\theequation}{S\arabic{equation}}
        \setcounter{section}{0}
        \renewcommand{\thesection}{S\arabic{section}}
        \setcounter{table}{0}
        \renewcommand{\thetable}{S\arabic{table}}%
        \setcounter{figure}{0}
        \renewcommand{\thefigure}{S\arabic{figure}}%
     }
\newcommand{\Exp}{\mathbb{E}}
\newcommand{\Cov}{{\rm Cov}}
\newcommand{\Var}{{\rm Var}}
\newcommand{\vecc}{{\rm vec}} 
\newcommand{\vech}{{\rm vech}} 
\newcommand{\diag}{{\rm diag}} 
\newcommand{\tildeK}{\widetilde{K}} 
\newcommand{\nablaa}{\widetilde{\nabla}} 
\newcommand{\Partials}{\if1\bolding{\widetilde{\bm{\partial}_\bs}}\fi\if0\bolding{\widetilde{\partial}_\bs}\fi}
\newcommand{\Sigmaa}{\widetilde{\Sigma}} 
\newcommand{\given}{\, | \;} 
\newcommand{\mstar}{m^*}
\def\A{\mathscr{A}}
\def\C{\mathscr{C}}
\def\G{\if1\bolding{\bm{\mathscr{G}}}\fi\if0\bolding{\mathscr{G}}\fi}
\def\I{\mathcal{I}}
\def\calK{\if1\bolding{\bm{\mathcal{K}}}\fi\if0\bolding{\mathcal{K}}\fi}
\def\L{\if1\bolding{\bm{\mathcal{L}}}\fi\if0\bolding{\mathcal{L}}\fi}
\def\M{\mathcal{M}}
\def\N{\mathcal{N}}
\def\P{\mathcal{P}}
\def\D{\mathcal{D}}
\def\Y{\if1\bolding{\bm{\mathcal{Y}}}\fi\if0\bolding{\mathcal{Y}}\fi}
\def\Z{\if1\bolding{\bm{\mathcal{Z}}}\fi\if0\bolding{\mathcal{Z}}\fi}
\def\R1{\mbox{$\mathfrak{R}$}}
\def\R2{\mbox{$\mathfrak{R}^2$}}
\def\R3{\mbox{$\mathfrak{R}^3$}}
\def\Rd{\mbox{$\mathfrak{R}^{d}$}}
\def\Rq{\mbox{$\mathfrak{R}^{q}$}}
\newtheorem{definition}{Definition}
\newtheorem{theorem}{Theorem}
\newtheorem{result}{Result}
\newtheorem{condition}{Condition}
\newtheorem{lemma}{Lemma}
\newtheorem{corollary}{Corollary}
\newtheorem{proposition}{Proposition}
\newtheorem{assumption}{Assumption}
\newtheorem{remark}{Remark}
\def\bnabla{\if1\bolding{\bm{\nabla}}\fi\if0\bolding{\nabla}\fi}
\def\bpartial{\if1\bolding{\bm{\partial}}\fi\if0\bolding{\partial}\fi}
\def\0{\if1\bolding{\bm{0}}\fi\if0\bolding{0}\fi}
\def\1{\if1\bolding{\bm{1}}\fi\if0\bolding{1}\fi}
\def\ba{\if1\bolding{\bm{a}}\fi\if0\bolding{a}\fi}
\def\be{\if1\bolding{\bm{e}}\fi\if0\bolding{e}\fi}
\def\bu{\if1\bolding{\bm{u}}\fi\if0\bolding{u}\fi}
\def\bv{\if1\bolding{\bm{v}}\fi\if0\bolding{v}\fi}
\def\bw{\if1\bolding{\bm{w}}\fi\if0\bolding{w}\fi}
\def\bW{\if1\bolding{\bm{W}}\fi\if0\bolding{W}\fi}
\def\bx{\if1\bolding{\bm{x}}\fi\if0\bolding{x}\fi}
\def\bs{\if1\bolding{\bm{s}}\fi\if0\bolding{s}\fi}
\def\bDelta{\if1\bolding{\bm{\Delta}}\fi\if0\bolding{\Delta}\fi}
\def\bP{\if1\bolding{\bm{\mathrm{P}}}\fi\if0\bolding{P}\fi} 
\def\bmu{\if1\bolding{\bm{\mu}}\fi\if0\bolding{\mu}\fi}
\def\btheta{\if1\bolding{\bm{\theta}}\fi\if0\bolding{\theta}\fi}
\def\bbeta{\if1\bolding{\bm{\beta}}\fi\if0\bolding{\beta}\fi}
\def\bgamma{\if1\bolding{\bm{\gamma}}\fi\if0\bolding{\gamma}\fi}
\def\bL{\if1\bolding{\bm{L}}\fi\if0\bolding{L}\fi}
\def\bV{\if1\bolding{\bm{\mathrm{V}}}\fi\if0\bolding{V}\fi} 
\def\bn{\if1\bolding{\bm{n}}\fi\if0\bolding{n}\fi}
\def\m{\if1\bolding{\bm{m}}\fi\if0\bolding{m}\fi}
\def\bH{\if1\bolding{\bm{\mathrm{H}}}\fi\if0\bolding{H}\fi} 
\def\bN{\if1\bolding{\bm{\mathrm{N}}}\fi\if0\bolding{N}\fi} 
\def\bGamma{\if1\bolding{\bm{\Gamma}}\fi\if0\bolding{\Gamma}\fi}
\def\bF{\if1\bolding{\bm{F}}\fi\if0\bolding{F}\fi}
\def\bA{\if1\bolding{\bm{\mathrm{A}}}\fi\if0\bolding{A}\fi} 
\def\bB{\if1\bolding{\bm{\mathrm{B}}}\fi\if0\bolding{B}\fi} 
\def\bK{\if1\bolding{\bm{\mathrm{K}}}\fi\if0\bolding{K}\fi} 
\def\bG{\if1\bolding{\bm{\mathrm{G}}}\fi\if0\bolding{G}\fi} 
\def\bI{\if1\bolding{\bm{\mathrm{I}}}\fi\if0\bolding{I}\fi} 
\def\bX{\if1\bolding{\bm{\mathrm{X}}}\fi\if0\bolding{X}\fi} 
\def\bR{\if1\bolding{\bm{\mathrm{R}}}\fi\if0\bolding{R}\fi} 
\def\bE{\if1\bolding{\bm{\mathrm{E}}}\fi\if0\bolding{E}\fi} 
\def\bO{\if1\bolding{\bm{\mathrm{O}}}\fi\if0\bolding{O}\fi} 
\def\bU{\if1\bolding{\bm{\mathrm{U}}}\fi\if0\bolding{U}\fi}
\def\bM{\if1\bolding{\bm{\mathrm{M}}}\fi\if0\bolding{M}\fi}
\maketitle

\begin{abstract}
    This paper develops Bayesian wombling, or boundary analysis for tracking invasive species. We focus on three invasive species of concern in North America: \emph{Lycorma delicatula}, \emph{Adelges tsugae}, and \emph{Microstegium vimineum}. We aim to demonstrate the applicability of wombling for supporting invasive pest management activities. In spatial science wombling is used widely to track regions of rapid change within the spatial reference domain. We use Gaussian process models for point-referenced spatial data followed by predictive inference on rates of change over curves using line integrals. We extend the wombling model based Bayesian inference framework to feature simple analytic forms that avoid high dimensional quadrature while offering uncertainty quantification over curvilinear spatial boundaries.
\end{abstract}

\section{Introduction}


\begin{figure}[t]
    \centering
    \includegraphics[width=\linewidth]{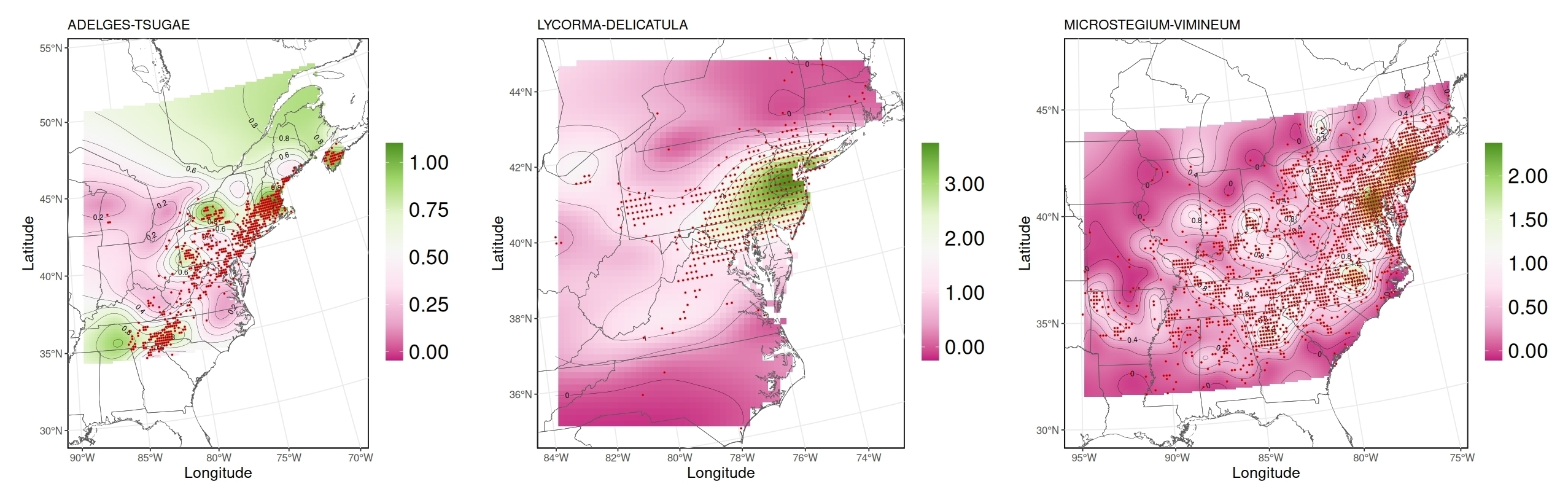}
    \caption{Interpolated surfaces for logarithm of species occurrence counts for (left) \emph{Adelges tsugae} (center) \emph{Lycorma delicatula} (right) \emph{Microstegium vimineum}}
    \label{fig:species}
\end{figure}

Invasive alien species can have substantial biodiversity, environmental and economic impacts, with global impacts of invasive alien species estimated at more than \$9 billion annually from 1970 - 2020 \citep[see, e.g.,][]{leroy2022analysing}. These impacts can be incredibly dependent on where the species invade, making species occurrence data and estimates of species distributions critical to being able to respond to, manage invasive species and devise interventions \citep[see, e.g.,][]{latzka2016spatialheterogeneity}. A common approach to managing invasive species applies surveillance based on where a species has been found, high risk areas of incursion, potential impacts, and species distribution models to guide efforts \citep[see, e.g.,][]{epanchinniell2010controlinvaders}. { In the case of species distribution models, these often estimate the potential area that a species could establish (i.e. potential niche; e.g. MAXENT \cite{horrocks2024biology}, CLIMEX \cite{taylor2012climate}, CLIMATCH \cite{bomford2008risk}, Bioclim \cite{panetta1987bioclimatic}), rather than its actual distribution (though see \cite{lindenmayer1991conservation} and \cite{vaclavik2009invasive} for examples of estimating actual distributions). \cite{vaclavik2009invasive} illustrate the importance of including constraints when trying to estimate actual species distributions. Therefore, estimating the boundaries of a species invasion would support effective management of invasive species under the typical conditions of resource constraints, by providing a more realistic determination of where to concentrate efforts.}

At present, species distribution, phenology, and species distribution models are used to support risk assessment, risk management, and surveillance activities for invasive species \citep[see, e.g.,][]{srivastava2019sdm,hudgins-2020,belouard-slf-jump2025}. These models generally provide a hypothesis of the potential range into which an invasive species can expand or create potential spread scenarios for planning management actions. While this information can guide surveillance and risk mitigation efforts, there is still considerable time spent determining optimal locations for pest survey sites to delimit and track the progress of an invasion, as well as determining whether a mitigation or management measure is effective at reducing the spread of a species. { Identifying regions of rapid change in occurrence remains crucial for generating a hypothesis of the actual, rather than potential, distribution of an invasive species, as well as identifying spatial transitions that may be associated with invasion boundaries, barriers, or corridors.} 

{ Detecting regions of rapid change in spatial distribution of a species and subsequently the effect or presence of boundaries, commonly  known as wombling \citep[see, e.g.][]{womble_differential_1951,gleyze2001wombling,banerjee2010handbook}, has been of ecological interest, for example, \cite{holway_edge_2005} examines effects across a natural boundary in Argentine ants using standard analysis of variance, \cite{fitzpatrick_ecological_2010} looks at boundaries in the spatial distribution of the hemlock woolly adelgid at the county-level, boundaries being shared regions for neighboring counties, for further examples \citep[see, e.g.][]{gao_spatial_2023,wu_assessing_2025}. Wombling for point-referenced data, for instance, latitude-longitude or, easting-northing referenced data, demands substantially different methods, for example, \cite{strydom_spatialboundariesjl_2023} considers formulating rates of change for woody plant distributions and looks for candidate boundaries only at sharp elevation changes, their spatial interpolation is deterministic. 
We are concerned with point-referenced Bayesian wombling under a hierarchical model \citep[see, e.g.,][]{banerjee_directional_2003,banerjee2006bayesian,qu_boundary_2021,halder_bayesian_2024} for the spatial distribution of an invasive species. Our choices of candidate boundaries are driven by high gradients in species distribution. Developing an easily-accessible framework that implements Bayesian wombling for use by the wider ecological community is faced with several challenges. Perhaps, the most severe being computational expense
. Our statistical contributions lie in the use of analytic closed forms for posteriors that improves upon previous approaches while producing }
efficient Bayesian inference for hierarchical spatial models \citep[see, e.g.][]{banerjee_hierarchical_2014} and the novel application to tracking invasive species. We apply Bayesian wombling to detect boundaries in the distribution of three invasive species of concern in North America: \emph{Lycorma delicatula} (LD) { commonly known as the spotted {\em lantern fly}}, \emph{Adelges tsugae} (AT) { commonly known as the {\em hemlock woolly adelgid}}, and \emph{Microstegium vimineum} (MV) { commonly known as the {\em Japanese stiltgrass}}, with the objective of demonstrating its applicability for supporting invasive pest {and species} management activities. { The first two are insects and the third one is a plant species.}

{ There are no standard statistical approaches that offer model-based inference over boundaries (curves) within a spatial setting. However, when regions of interest are in question commonly used approaches are clustering \citep[see, e.g.,][]{ester_density-based_1996} and hot-spot detection \citep[see, e.g,][]{bivand_comparing_2018}.  While wombling seeks predictive inference over curves offering fully model-based uncertainty quantification leveraging rates of change, clustering and hot-spot detection techniques aim to find homogeneous regions in the reference domain, unable to offer inference over curves separating such regions. We show detailed comparisons. We make note that the data considered here are spatial rather than spatiotemporal, these gradients characterize spatial changes in species occurrence and should not, by themselves, be interpreted as estimates of temporal spread. Rather, they identify spatial transitions that may be consistent with invasion fronts, barriers, corridors, or other features of the species distribution.}

The manuscript evolves as follows: \Cref{sec:spatial_derivatives} elaborates on the processes that govern spatial rates of change followed by \Cref{sec:bayesian_wombling} which describes the underpinnings of Bayesian wombling and introduces the necessary results for computational improvements. \Cref{sec:review_bhm} reviews Bayesian hierarchical models and posterior inference. \Cref{sec:sim} calibrates the performance of Bayesian wombling with closed forms on simulated data. \Cref{sec:inv_species} { demonstrates the application to modeling invasive species counts, analyzing spatial rates of change in occurrence, and detecting boundaries.} \Cref{sec:diss} concludes with a discussion and future avenues of research.


\section{Spatial Derivative Processes}\label{sec:spatial_derivatives}
We consider $\{Y(\bs): \bs\in\mathscr{S}\subseteq\mathfrak{R}^2\}$ to be a univariate weakly stationary random field with zero mean and a positive definite covariance $K(\bs, \bs')= \Cov(Y(\bs), Y(\bs'))$ for locations $\bs, \bs'\in \mathfrak{R}^2$. Mean square smoothness \citep[see, e.g.,][]{stein_interpolation_1999} at an arbitrary location $\bs_0$ requires $Y(\bs_0+h\bu) = Y(\bs_0)+h\bu^{\T}\bpartial Y(\bs_0) + h^2 {\bu^{\otimes 2}}^{\T}\bpartial^{\otimes 2}Y(\bs_0) + o(h^3||\bu||^3)$, where, $\bu=(u_1,u_2)^{\T}\in \mathfrak{R}^2$ is an arbitrary vector of directions, $\bpartial$ is the gradient operator, $\bpartial Y(\bs_0) = \left(\frac{\partial}{\partial s_x}Y(\bs_0),\frac{\partial}{\partial s_y}Y(\bs_0)\right)^{\T}=(\partial_xY(\bs_0), \partial_yY(\bs_0))^{\T}$, $\otimes$ is the Kronecker vector product. Hence, $\bu^{\otimes 2}= \left(u_1^2, u_1u_2, u_2u_1, u_2^2\right)^{\T}$ and $\bpartial^{\otimes 2} = \bpartial \otimes \bpartial$. Note that $\bpartial^{\otimes 2} Y(\bs_0)$ is the vectorized Hessian. The processes $\bpartial Y(\bs_0)$ and $\bpartial^{\otimes 2}Y(\bs_0)$ govern rates of change in $Y(\bs_0)$. The \emph{gradient} or, first order rate of change, is captured by $\bpartial Y(\bs_0)$ while, \emph{curvature} is captured by $\bpartial^{\otimes 2}Y(\bs_0)$. \emph{Mean square differentiability} \citep[see][]{banerjee_smoothness_2003} of the first and second order for $Y(\bs_0)$ guarantees that $\bu^{\T}\bpartial Y(\bs_0)$ and $(\bu\otimes\bv)^{\T}\bpartial^{\otimes 2}Y(\bs_0)$ are well-defined respectively, for any set of direction vectors $\bu, \bv\in \mathfrak{R}^2$. We note that the entries of $\bpartial^{\otimes 2}Y(\bs_0)$ contain duplicates, both $\partial^2_{xy}Y(\bs_0)=\frac{\partial^2}{\partial s_x\partial s_y}Y(\bs_0)$ and $\partial^2_{yx}Y(\bs_0)$ are included. To avoid singularities that arise from duplication, we work with $\widetilde{\bpartial}^{\otimes 2}Y(\bs_0)=\left(\partial_{xx}^2Y(\bs_0),\partial_{xy}^2Y(\bs_0), \partial_{yy}^2Y(\bs_0)\right)^{\T}$ comprised of only unique derivatives.

Statistical inference is devised for the joint process, $\L^*Y(\bs) = \left(Y(\bs), \bpartial Y(\bs)^{\T}, \widetilde{\bpartial}^{\otimes 2}Y(\bs)^{\T}\right)^{\T}$. Validity of the inference is considered at length in 
\cite{banerjee_directional_2003,halder_bayesian_2024}. The process $\L^*Y(\bs)$ is also weakly stationary with a cross-covariance matrix, 

\begin{equation}\label{eq:cross-cov}
    \bV_{\L^*}(\bDelta) =\left(\begin{array}{ccc}
        K(\bDelta) &  \bpartial K(\bDelta)^{\T} & \widetilde{\bpartial}^2 K(\bDelta)^{\T} \\
         -\bpartial K(\bDelta) & -\bpartial^2K(\bDelta) & -\widetilde{\bpartial}^3K(\bDelta)^{\T}\\
         \widetilde{\bpartial}^{2}K(\bDelta) & \widetilde{\bpartial}^3K(\bDelta) & \widetilde{\bpartial}^4K(\bDelta)
    \end{array}\right), 
\end{equation}
where $\bDelta = \bs-\bs'$, $\bpartial K(\bDelta)$ is a $2\times 1$ vector of gradients, $\widetilde{\bpartial}^2K(\bDelta)$ is a $3\times 1$ vector of unique curvatures, $\widetilde{\bpartial}^3K(\bDelta)$ is a $3\times 2$ matrix of third derivatives, $\bpartial^2K(\bDelta)$ is the $2\times 2$ Hessian and $\widetilde{\bpartial}^4K(\bDelta)$ is a $3\times 3$ matrix of fourth order derivatives. Evidently, for $\bV_{\L^*}(\bDelta)$ in \cref{eq:cross-cov} to be valid all entries need to be well-defined.

Let $Y(\bs)\sim GP(0, K(\cdot;\btheta))$ denote a Gaussian process (GP) where $K(\bDelta;\btheta)= \Cov(Y(\bs), Y(\bs'))$ with process parameters $\btheta = \{\sigma^2, \phi\}$. We will denote $K(\bDelta;\btheta)=K(\bDelta)$ to ease notation. The covariance function satisfies $\sum_{i=1}^{N}\sum_{j=1}^{N}a_ia_jK(\bDelta_{ij})>0$ for any collection of coordinates $\{\bs_i:i = 1,\ldots,N\}$. Under isotropy we have $K(\bDelta)=\widetilde{K}(||\bDelta||)$. Let $\Y = \left(y(\bs_1), \ldots, y(\bs_N)\right)^{\T}$ be the observed realization over $\mathscr{S}$, $\Sigma_\Y$ be the $N\times N$ covariance matrix with entries $K(\bs_i,\bs_j)$, $i, j = 1, \ldots, N$ and $\bs_0$ an arbitrary location of interest. The joint distribution is as follows:

\begin{equation}\label{eq:joint}
   \Y, \bpartial Y(\bs_0), \widetilde{\bpartial}^2Y(\bs_0)\given \btheta\sim  \N_{N+5}\left(\0_{N+5},\left(\begin{array}{ccc}
         \Sigma_\Y & \bK_1 & \bK_2\\
         -\bK_1^{\T} & -\bpartial^2 K(\0) & -\widetilde{\bpartial}^3K(\0)^{\T}\\
         \bK_2^{\T} & \widetilde{\bpartial}^3K(\0) & \widetilde{\bpartial}^4K(\0)
    \end{array}\right)\right),
\end{equation}
where, $\N_d$ denotes the $d$-variate Gaussian distribution, $\bK_1 = \left(\bpartial K(\delta_{10})^{\T},\ldots,\bpartial K(\delta_{N0})^{\T}\right)^{\T}$, $\bK_2 = \left(\widetilde{\bpartial}^2 K(\delta_{10})^{\T},\ldots,\widetilde{\bpartial}^2 K(\delta_{N0})^{\T}\right)^{\T}$ and $\delta_{i0} =\bs_i-\bs_0$, $i = 1, \ldots, N$. The posterior predictive distribution for rates of change at $\bs_0$ is  $P(\bpartial Y(\bs_0),\widetilde{\bpartial}^2Y(\bs_0)\given \Y) = \int P(\bpartial Y(\bs_0),\widetilde{\bpartial}^2Y(\bs_0)\given \Y,\btheta)\; P(\btheta\given \Y)\;d\btheta$. Posterior sampling proceeds in a one-for-one fashion corresponding to posterior samples of $\btheta$. From \cref{eq:joint} the resulting full conditional distribution is obtained as follows:
\begin{equation}\label{eq:posterior_gradient}
    \left(\begin{array}{c}
         \bpartial Y(\bs_0) \\
         \widetilde{\bpartial}^2Y(\bs_0) 
    \end{array}\right)\given \Y\sim\N_5\left(-\left(\begin{array}{c}
         \bK_1 \\
         \bK_2 
    \end{array}\right)^{\T}\Sigma_\Y^{-1}\Y, \left(\begin{array}{cc}
         -\bpartial^2 K(\0) & -\widetilde{\bpartial}^3K(\0)^{\T} \\
         \widetilde{\bpartial}^3K(\0) & \widetilde{\bpartial}^4K(\0) 
    \end{array}\right)-\left(\begin{array}{c}
         \bK_1 \\
         \bK_2 
    \end{array}\right)^{\T}\Sigma_\Y^{-1}\left(\begin{array}{c}
         \bK_1 \\
         \bK_2 
    \end{array}\right)\right).
\end{equation}
We use the Mat\'ern class of kernels, $K(||\bDelta||,\btheta) = \sigma^2\Gamma(\nu)^{-1}2^{1-\nu}\left(\sqrt{2\nu}\phi||\bDelta||\right)^\nu K_\nu\left(\sqrt{2\nu}\phi||\bDelta||\right)$, where $K_\nu(\cdot)$ is the modified Bessel function of the second kind \cite[][]{abramowitz1988handbook} featuring a fractal parameter $\nu$ that controls process smoothness, a spatial range parameter $\phi$ and an overall variance parameter $\sigma^2$. { Note that the GP needs to be sufficiently smooth to ensure the existence of all differential processes, specifically, $\nu>2$ ensures the existence of derivative and the curvature process.}

\section{Bayesian Wombling}\label{sec:bayesian_wombling}
{ In general wombling seeks posterior predictive inference on line integrals over curves
\begin{equation}\label{eq:wombling_measures}
    \bGamma(C) = \left(\int_C \bu^{\T}\bpartial Y(\bs)\;d\bs, \int_C{\bu^{\otimes 2}}^{\T}\bpartial^{\otimes 2} Y(\bs)\;d\bs\right)^{\T},
\end{equation}
where $C$ is a curve of interest to the investigator \citep[see, e.g.,][Sec. 3]{halder_bayesian_2024, banerjee2006bayesian}. A curve, $C$ is said to form a {\em wombling boundary} if the total, or average wombling measures, which are defined as $\overline{\bGamma}(C) = \bGamma(C)/\ell(C)$, where $\ell$ is the arc-length measure, are significantly large. The curves can be topologically open or closed, we use $\int$ or $\oint$ accordingly in \cref{eq:wombling_measures}. Selecting an appropriate direction, $\bu$ becomes crucial when measuring rates of change. The curve $C$ typically tracks a region of rapid change in the spatial domain and hence, the direction normal to $C$ is of importance. We denote the normal to $C$ at $\bs$ by $\bn(\bs)$ and set $\bu = \bn(\bs)$ in the line integrals of \cref{eq:wombling_measures}. The curve $C$ is deemed to be a \emph{wombling boundary} with regard to the gradient (and/or curvature) if the first (and/or second) element of $\bGamma(C)$ is large. Focusing on the choices for $C$, clearly not all curves ensure the existence of $\bn(\bs)$ at every $\bs$. Parametric smooth curves offer some respite in that regard. We work with $C = \{\bs(t) = (s_1(t), s_2(t)):t \in\I \subset \mathfrak{R}\}$. As $t$ varies over $\I$, $\bs(t)$ traces out $C$. We assume $||\bs'(t)||\ne0$ which ensures $\bn(\bs)=||\bs'(t)||^{-1}\left(s_2'(t), -s_1'(t)\right)^{\T}$ is well-defined.
}

The arc-length, $\ell(C) = \int_\I||\bs'(t)||\;dt$. For parametric curves $\bGamma(C)$ can be expressed as, $\bGamma(C) = \left(\int_\I \bn(\bs(t))^{\T}\bpartial Y(\bs(t))||\bs'(t)||\;dt, \int_\I{\bn(s(t))^{\otimes 2}}^{\T}\bpartial^{\otimes 2} Y(\bs(t))||\bs'(t)||\;dt\right)^{\T}$. Let $\I = [0,t^*]$, and the curve traced out over $\I$ be denoted as $C_{t^*}$. Statistical inference for $\bGamma(C_{t^*})$ follows from $\bpartial Y(\bs)$ and $\widetilde{\bpartial}^2Y(\bs)$ being GPs, as seen in \cref{eq:posterior_gradient}, $\bGamma(C_{t^*})\sim\N_2(\0_2, \bK_\bGamma(t^*,t^*))$, where $\bK_\bGamma(t^*,t^*)$ is a $2\times 2$ matrix with entries 
\begin{equation}\label{eq:variance_wm}
    k_{ij}(t^*,t^*)= (-1)^i\int_0^{t^*}\int_0^{t^*}\ba_i(t_1)^{\T}\;\bpartial^{i+j}K(\bDelta(t_1,t_2))\;\ba_j(t_2)\;||\bs'(t_1)||\;||\bs'(t_2)||\;dt_1\;dt_2,
\end{equation}
where $\ba_1(t)=\bn(\bs(t))$, $\ba_2(t)=\mathcal{E}_2\;\bn(\bs(t))^{\otimes 2}$, with $\mathcal{E}_2= \left(\begin{smallmatrix}
        1 & & & \\ & 1 & 1 & \\ & & & 1
    \end{smallmatrix}\right)$ being an elimination matrix and { $\bDelta(t_1,t_2)=\bs(t_2)-\bs(t_1)$} for $i,j = 1,2$. Predictive inference on $\bGamma(C)$ requires
\begin{equation}\label{eq:joint-wm}
   \Y, \bGamma(C_{t^*})\given \btheta\sim  \N_{N+2}\left[\0_{N+2},\left(\begin{array}{cc}
         \Sigma_\Y & \G_\bGamma(t^*)^{\T} \\
         \G_\bGamma(t^*) & \bK_\bGamma(t^*, t^*)
    \end{array}\right)\right],
\end{equation}
where $\G_\bGamma(t^*)^{\T}=\left(\begin{smallmatrix}\bgamma_1(t^*)^{\T} \\ \vdots \\\bgamma_N(t^*)^{\T}\end{smallmatrix}\right)$ is an $N\times 2$ matrix with entries 
\begin{equation}\label{eq:cross-cov-wm}
    \gamma_i(t^*)=\left(\int_0^{t^*}\bn(\bs(t))^{\T} \bpartial K(\bDelta_j(t))\;||\bs'(t)||\;dt, \int_0^{t^*}{\bn(\bs(t))^{\otimes 2}}^{\T} \bpartial^{\otimes 2} K(\bDelta_j(t))\;||\bs'(t)||\;dt\right)^{\T},
\end{equation}
where $\bDelta_j(t) =  \bs(t) - \bs_j$. Posterior predictive inference proceeds one-for-one (similar to \cref{eq:posterior_gradient}) using $\bGamma(C_{t^*})\given \Y\sim \N_2\left(-\G_\bGamma(t^*)\Sigma_\Y^{-1}\Y, \bK_\bGamma(t^*, t^*) - \G_\bGamma(t^*)\Sigma_\Y^{-1}\G_\bGamma(t^*)^{\T}\right)$.

In practice, modern computing environments store curves as a set of points. As a result, it suffices to demonstrate wombling for rectilinear approximations to smooth curves where predictive inference is performed iteratively on segments. We show the inference for one generic segment. Let $C =\{\bs(t_0), \bs(t_1), \ldots, \bs(t_{n_p})\}$, then the $i$-th segment, $C_{t_i}=\{\bs(t) = \bs(t_{i-1})+t\bu_i:t\in[0,t_i]\}$, $t_i = ||\bs(t_i)-\bs(t_{i-1})||$ and $\bu_i = t_i^{-1}(\bs(t_i)-\bs(t_{i-1}))$. Clearly, $||\bu_i||=1$, $||\bs'(t)||=1$ and the normal to $C_{t_i}$ is $\bu_i^{\perp}=(u_{i2},-u_{i1})^{\T}$. For predictive inference on $\bGamma(C_{t_i})$ we set $\bDelta(t_1, t_2)=(t_2-t_1)\bu_i$ in \cref{eq:variance_wm} and $\bDelta_j(t)=\bDelta_{i-1,j}+t\bu_i=(\bs_{i-1}-\bs_j) + t\bu_{i-1}$ in \cref{eq:cross-cov-wm}.

\subsection{Closed Forms}
We acknowledge that \cref{eq:variance_wm} requires 2-dimensional quadrature which is computationally expensive to evaluate. { Conventional approaches to wombling \citep[see, e.g.][]{banerjee2006bayesian,halder_bayesian_2024} evaluate \cref{eq:variance_wm} using two-dimensional quadrature using a  Mat\'ern kernel. We improve on this by showing that closed form analytic expressions exist for the entries of $\bK_\bGamma(t^*,t^*)$. The benefits are reduced computation time and ease of implementation for which specific details are in \Cref{sec:sim}}. The following results provide more details. 
\begin{theorem}\label{th:1}
         On each rectilinear segment, $C_{t^*}=\{\bs_0+t\bu:t\in[0,t^*]\}$, where $\bu$ is a unit vector and $\bu^\perp$ is its normal, considering the terms in the cross-covariance matrix we have
         
         \begin{equation*}
             \begin{split}
                 k_{ij}(t^*,t^*)&= (-1)^i\int_0^{t^*}\int_0^{t^*}\ba_i(t_1)^{\T}\;\bpartial^{i+j}K(\bDelta(t_1,t_2))\;\ba_j(t_2)\;||\bs'(t_1)||\;||\bs'(t_2)||\;dt_1\;dt_2,\\
                 &=(-1)^i\int_{-t^*}^{t^*}(t^*-|x|)\;\ba_i^{\T}\;\bpartial^{i+j}K(x\bu)\;\ba_j\;dx, \quad i,j = 1,2,
             \end{split}
         \end{equation*}
         where $\ba_1(t)=\bn(\bs(t))$ is the normal to the segment, $\ba_2(t)=\mathcal{E}_2\;\bn(\bs(t))\otimes \bn(\bs(t))$, where $\mathcal{E}_2= \left(\begin{smallmatrix}
        1 & & & \\ & 1 & 1 & \\ & & & 1
    \end{smallmatrix}\right)$ is an elimination matrix and { $\bDelta(t_1,t_2)=\bs(t_2)-\bs(t_1)$}.
\end{theorem}
    \begin{proof}
        Considering the parametric segment, $C_{t^*}=\{\bs_0+t\bu:t\in[0,t^*]\}$, where $\bu=(u_1,u_2)^{\T}$ is a unit vector, $||\bu||=1$, $\bu^\perp$ is its normal, $\bu^{\T}\;\bu^\perp=0$, $\ba_1(t)=\ba_1=\bu^\perp$ and $\ba_2(t)=\ba_2=\mathcal{E}_2(\bu^\perp\otimes \bu^\perp)$ are free of $t$, and $\bDelta(t_1,t_2)=(t_2-t_1)\bu$. The integrand in $k_{ij}(t^*,t^*)= (-1)^i\int_0^{t^*}\int_0^{t^*}\ba_i^{\T}\;\bpartial^{i+j}K((t_2-t_1)\bu)\;\ba_j\;dt_1\;dt_2$, depends only on $(t_2-t_1)$. Making a change of variable--define $x=t_2-t_1$, $y=t_2+t_1$, implying $\frac{x+y}{2}=t_2$ and $\frac{y-x}{2}=t_1$. The Jacobian is $\frac{1}{2}$. Hence, $0\leq y+x\leq 2t^*$ and $0\leq y-x\leq 2t^*$. This implies $|x| \leq y \leq 2t^*-|x|$ and $-t^*\leq x \leq t^*$. Making the substitution above reduces, $k_{ij}(t^*,t^*)= \frac{(-1)^i}{2}\int_{|x|}^{2t^*-|x|}\int_{-t^*}^{t^*}\ba_i^{\T}\;\bpartial^{i+j}K(x\bu)\;\ba_j\;dx\;dy=(-1)^i\int_{-t^*}^{t^*}(t^*-|x|)\;\ba_i^{\T}\;\bpartial^{i+j}K(x\bu)\;\ba_j\;dx$. Setting $i=j=1$, produces the scenario in \cite{banerjee2006bayesian}. 
    \end{proof}

    { The Mat\'ern covariance has analytic expressions for specific values of the fractal parameter, 
    \begin{align}\label{eq:cov-nu}
    	K(||\bDelta||,\btheta)=\begin{cases}
    		\sigma^2(1+\sqrt{3}\phi||\bDelta||)e^{-\sqrt{3}\phi||\bDelta||}&\nu=3/2\\
    		\sigma^2\left(1+\sqrt{5}\phi||\bDelta||+\mfrac{5}{3}\phi^2||\bDelta||^2\right)e^{-\sqrt{5}\phi||\bDelta||}& \nu=5/2
    	\end{cases}.
    \end{align}
    For $\nu=3/2$ it is once mean square differentiable and for $\nu=5/2$ it is twice mean square differentiable at 0.
    }

    \begin{theorem}\label{th:2}
        For the Mat\'ern kernel, with $\nu = 3/2$ in \cref{eq:cov-nu}, the variance of the gradient based wombling measure is  $k_{11}(t^*,t^*)=2\sigma^2\left(\sqrt{3}\phi t^*-1+e^{-\sqrt{3}\phi t^*}\right)$. In  case $\nu=5/2$ in \cref{eq:cov-nu}, the cross-covariance matrix for the wombling measures based on spatial gradient and curvature requires $k_{11}(t^*,t^*)=\frac{2\sigma^2}{3} \left(2\sqrt{5}\phi\;t^*-3+(\sqrt{5}\phi t^*+3)e^{-\sqrt{5}\phi t^*}\right)$, $k_{22}(t^*,t^*)=10\sigma^2\phi^2\left(\sqrt{5}\phi t^*-1+e^{-\sqrt{5}\phi t^*}\right)$ and $k_{21}(t^*,t^*)=-k_{12}(t^*,t^*)=0$. For the Gaussian kernel, 

        \begin{equation*}
            \bV_{\bGamma}(t^*)=2\sigma^2\left[\sqrt{\pi}\phi t^*\left\{2\Phi\left(\sqrt{2}\phi t^*\right)-1\right\}-(1-e^{-\phi^2{t^*}^2})\right]\left(\begin{smallmatrix} 1 & 0\\0&6\phi^2 \end{smallmatrix}\right),
        \end{equation*}
        where $\Phi(\cdot)$ is the cdf for the standard Gaussian probability density.
    \end{theorem}
    \begin{proof}
        Using \Cref{th:1}, if $\nu = 3/2$, then we obtain $k_{11}(t^*,t^*)=3\sigma^2\phi ^2\int_{-t^*}^{t^*}(t^*-|x|)e^{-\sqrt{3}\phi  |x|}\;dx=2\sigma^2\left(\sqrt{3}\phi t^*-1+e^{-\sqrt{3}\phi t^*}\right)$. If $\nu=5/2$, then $k_{11}(t^*,t^*)=\frac{5}{3}\sigma^2\phi\int_{-t^*}^{t^*}(t^*-|x|)(1+\sqrt{5}\phi |x|)\;e^{-\sqrt{5}\phi  |x|}\;dx=\frac{2\sigma^2}{3} \left(2\sqrt{5}\phi\;t^*-3+(\sqrt{5}\phi t^*+3)e^{-\sqrt{5}\phi t^*}\right)$. For terms $k_{21}$ and $k_{12}$, let us consider $a_1^{\T}\bpartial^3K(x\bu)\;a_2=\frac{25\sqrt{5}}{3}\sigma^2\phi^3 e^{-\sqrt{5}\phi  |x|}|x|\big\{\ba_1^{\T} \bA_1 \ba_2 -\sqrt{5}\phi |x|\; \ba_1^{\T}\; \bA_2 \;\ba_2\big\}$, where 
        
        \begin{equation*}
            \bA_1 = \left(\begin{smallmatrix} 3u_1 & u_2 & u_1\\u_2 & u_1 & 3u_2 \end{smallmatrix}\right), \quad \bA_2=\left(\begin{smallmatrix} u_1^3 & u_1^2\;u_2 & u_2^2\;u_1\\ u_1^2\;u_2 & u_2^2\;u_1 & u_2^3 \end{smallmatrix}\right).
        \end{equation*} 
        The first and the second term both reduce to 0. Hence, $k_{21}(t^*,t^*)=-k_{12}(t^*,t^*)=0$. Next, $\ba_2^{\T} \bpartial^4K(x\bu)\;\ba_2 = \frac{25}{3}\sigma^2\phi ^4\;e^{-\sqrt{5}\phi  |x|}\big\{\ba_2^{\T} \bA_3\; \ba_2 - \sqrt{5}\phi \; |x|\; \ba_2^{\T} \bA_4\Big(\sqrt{5}\;\phi \;|x|+1\Big) \;\ba_2\big\}$, where 
        
        \begin{equation*}
            \bA_3=\left(\begin{smallmatrix} 3 & 0 & 1\\0 & 1 & 0\\1 & 0 & 3 \end{smallmatrix}\right), \quad \bA_4(x')=\left(\begin{smallmatrix}
        6\;u_1^2-\;x'\; u_1^4 & (3-x'\; u_1^2)\;u_1\;u_2 & 1-x'\;u_1^2\;u_2^2\\ (3-x'\; u_1^2)\;u_1\;u_2 & 1-x'\;u_1^2\;u_2^2 & (3-x'\;u_2^2)\;u_1\;u_2\\1-x'\;u_1^2\;u_2^2 & (3-x'\;u_2^2)\;u_1\;u_2 & 6\;u_2^2-\;x'\; u_2^4
    \end{smallmatrix}\right).
        \end{equation*}
        After some algebra, $\ba_2^{\T}\; \bpartial^4K(x\bu)\;\ba_2=25\sigma^2\phi ^4\;e^{-\sqrt{5}\phi  |x|}\big\{1-2\sqrt{5}\;\phi \;|x|\;\bu_1^\perp \;\bu_2^\perp\;(\bu_1\;\bu_2+\bu_1^\perp\; \bu_2^\perp)\big\}$. Observe that $\bu_1^\perp = \bu_2$ and $\bu_2^\perp=-\bu_1$, substituting we get $k_{22}(t^*,t^*)=10\sigma^2\phi^2\left(\sqrt{5}\phi t^*-1+e^{-\sqrt{5}\phi t^*}\right)$. 
        
        For the Gaussian kernel following the same steps we obtain,
    $k_{11}(t^*,t^*)=2\sigma^2\phi^2 \int_{-t^*}^{t^*}(t^*-|x|)e^{-\phi^2  x^2}\;dx=2\sigma^2\left[\sqrt{\pi}\phi t^*\left\{2\Phi\left(\sqrt{2}\phi t^*\right)-1\right\}-(1-e^{-\phi^2{t^*}^2})\right]$. 
    Note that, similar to before, $\ba_1^{\T}\bpartial^3K(x\bu)\;\ba_2=4\sigma^2\phi ^4\;e^{-\phi^2  x^2}x\left\{\ba_1^{\T} \bA_1 \ba_2 -2\phi^2  x^2\; \ba_1^{\T}\; \bA_2 \;\ba_2\right\}$, the first and second terms equate to 0 implying $k_{21}(t^*,t^*)=-k_{12}(t^*,t^*)=0$. Next, $\ba_2^{\T} \bpartial^4K(x\bu)\;\ba_2 = 4\sigma^2\phi^4\;e^{-\phi^2 x^2}\big\{\ba_2^{\T} \bA_3\; \ba_2 - 2\phi^2 \; x^2\; \ba_2^{\T} \bA_4(2\phi^2 x^2) \;\ba_2\big\}$ yielding, $\ba_2^{\T}\; \bpartial^4K(x\bu)\;\ba_2=12\sigma^2\phi^4\;e^{-\phi^2  x^2}\big\{1-4\;\phi^2 \;x^2\;\bu_1^\perp\; \bu_2^\perp\;(\bu_1\;\bu_2+\bu_1^\perp\;\bu_2^\perp)\big\}$. 
    Substituting, we get $k_{22}(t^*,t^*)=6\phi^2k_{11}$.
    \end{proof}
    \Cref{th:2} interestingly shows that the posited cross-covariance matrix, $\bV_{\bGamma}(t^*)$, reduces to a \emph{variance-covariance matrix}. { Note that \Cref{th:1,th:2} are valid only for $\nu=3/2$, $\nu=5/2$ and $\nu\to\infty$ (the squared exponential) kernel.} Compared with a Mat\'ern kernel featuring $\nu = \frac{5}{2}$, the matrix has a simpler form for the Gaussian case. Finally, considering the covariance between $Z(\bs_i)$, $i=1,\ldots, N$ and the wombling measures---in the Gaussian case, closed-forms are available \citep[see, e.g.,][end of Section 3]{halder_bayesian_2024}. The inferential exercise of wombling does not require quadrature when using a Gaussian kernel however, only one-dimensional quadrature is required for the same when using a Mat\'ern kernel.

\section{Bayesian Hierarchical Models}\label{sec:review_bhm}
    A Bayesian hierarchical model is specified as follows:
    \begin{equation}\label{eq:bhm}
        Y(\bs) = \mu(\bs, \bbeta)+Z(\bs)+\epsilon(\bs),
    \end{equation}
    where $\mu(\bs,\bbeta)=\bx(\bs)^{\T}\bbeta$, $Z(\bs)\sim GP(0,K(\cdot;\sigma^2,\phi))$ and $\epsilon(\bs)\sim N(0,\tau^2)$ is a white noise process. The process parameters are $\btheta =\{\sigma^2, \phi, \tau^2\}$. A customary \emph{collapsed} posterior \citep[see, e.g.][]{finley2019efficient} for $\btheta$ is specified as follows:
    \begin{equation}\label{eq:full_posterior}
        P(\btheta\given\Y)\propto U(\phi\given a_\phi,b_\phi)\times IG(\sigma^2\given a_\sigma,b_\sigma)\times IG(\tau^2\given a_\tau,b_\tau)\times \N_N(\Y\given \bX\bbeta, \Sigma +\tau^2\bI_N),
    \end{equation}
    where $\Sigma = \sigma^2\bR_\Z(\phi)$, with $\bR_\Z(\phi)$ being the correlation matrix corresponding to $K(\cdot;\sigma^2,\phi)$, $U(\cdot\given)$ is the uniform distribution and $IG(\cdot\given)$ is the inverse-gamma distribution. Hyper-parameters are chosen such that the resulting prior on $\btheta$ { has a large variance, being ``non-informative" in that sense. For example, common choices for $a_\phi=10^{-3}$ and $b_\phi=300$, where $a_\phi$ can be reduced and $b_\phi$ can be increased if the posterior estimate of $\phi$ from initial pilot runs is close to either one; $a_\sigma=b_\sigma=a_\tau=b_\tau=2$ so as to generate an IG distribution with mean of 2 and infinite variance.} Posterior samples for { $\Z=(Z(\bs_1),\ldots, Z(\bs_n))^\T$} and $\bbeta$ are generated one-for-one corresponding to posterior samples of $\btheta$ using a Gibbs sampling scheme. { For more details on the conditional distribution, $P(\btheta\given \Y)$ the interested reader is referred to \cite{halder_bayesian_2024}, Section 4.}

    {
    Note that, our discussion on differential processes has been solely on the process, $Z(\bs)$, these processes can be associated with $Y(\bs)$, provided $\mu(\bs,\bbeta)$ admits spatial smoothness and, the location at which we are interested in obtaining inference on gradients is not one at which the process is observed. Predictive inference is sought for $\L^*Z(\bs_0)$ at an arbitrary location $\bs_0$ through 
    \begin{equation}\label{eq:inference_gradients}
        P(\bpartial Z(\bs_0)^{\T}, \widetilde{\bpartial}^2Z(\bs_0)^{\T}\given \Y)=\int P(\bpartial Z(\bs_0)^{\T}, \widetilde{\bpartial}^2Z(\bs_0)^{\T}\given \Z,\btheta)\; P(\Z\given \Y,\btheta)\; P(\btheta\given \Y)\;d\btheta\;d\Z.
    \end{equation}
    }
    Similarly, for the wombling measures, inference proceeds using 
    \begin{equation}\label{eq:inference_wms}
        P(\bGamma_\Z(C_{t^*})\given \Y)=\int P(\bGamma_\Z(C_{t^*})\given \Z,\btheta)\;P(\Z\given \Y,\btheta)\;P(\btheta\given \Y)\;d\btheta\;d\Z.
    \end{equation}
    \begin{figure}[t]
    \centering
    \includegraphics[width=0.8\linewidth]{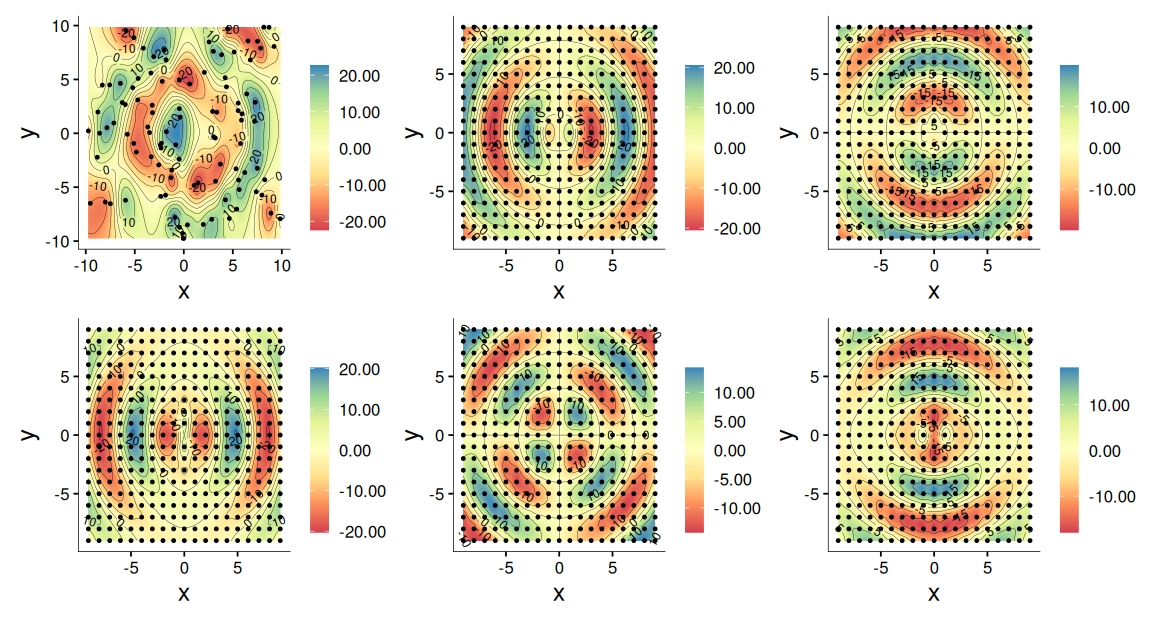}
    \includegraphics[width=0.8\linewidth]{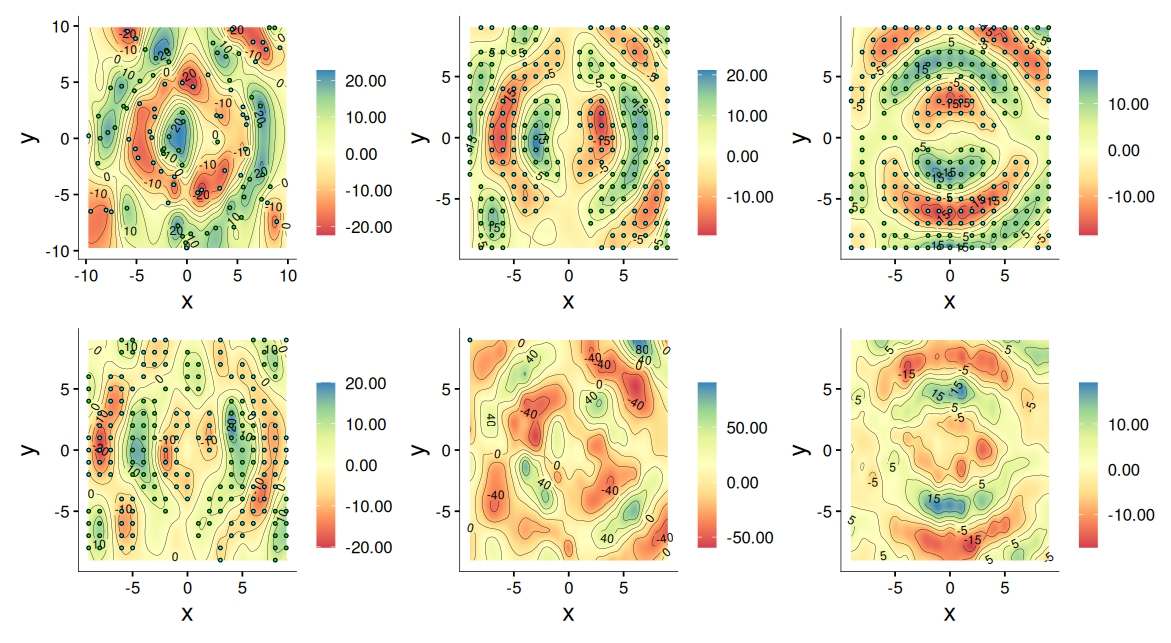}
    \caption{Interpolated surfaces for (first row: left) data used in simulation experiments (first row: center) true gradient along $x$ (first row: right) true gradient along $y$; (second row: left) true curvature along $x$ (second row: center) true mixed curvature (second row: right) true curvature along $y$; (third row: left) fitted data (third row: center) estimated gradient along $x$ (third row: right) estimated gradient along $y$; (fourth row: left) estimated curvature along $x$ (fourth row: center) estimated mixed curvature (fourth row: right) estimated curvature along $y$.}
    \label{fig:sim-est}
    \end{figure}

    \section{Simulation Experiments}\label{sec:sim}
    
    Gradients and curvatures are not observed in reality rather they are induced by a parent spatial process. For the purpose of calibrating our inference, we use data that is not generated by the model in \cref{eq:bhm} in our simulation experiments. We produce data using patterns that yield closed form expressions for rates of change. We generate $N=100$ observations over $[-10, 10] \times[-10, 10]\subseteq \mathfrak{R}^2$ arising from, $y(\bs)\sim \N_1\left(\mu_0(\bs) = 20\sin||\bs||, \tau^2\right)$, setting $\tau^2 = 1$. The patterned data is shown in the left plot, first row of \Cref{fig:sim-est}. The true values of gradients are available in closed-form. For example, $\partial_x \mu_0(\bs_G)=20\cos||\bs_G||\;s_{x,G}\;||\bs_G||^{-1}$, where $\bs_G =(s_{x,G}, s_{y,G})$ lies on a grid overlaid on the domain of reference (in this case: $[-10,10]\times[-10, 10]$). Other gradient and curvature processes are computed similarly by differentiating $\mu_0(\bs)$. The resulting plots are shown in the remaining plots within the first and second rows of \Cref{fig:sim-est}. We fit the model in \cref{eq:bhm} to the simulated data using only an intercept, letting $\Z$ learn the spatial variation.
    
    We fit the model in \cref{eq:bhm} using a Mat\'ern kernel with $\nu=\frac{5}{2}$ to the simulated data. This allows for inference on gradients and curvatures. Posterior samples of $\bbeta$, $\btheta$ are generated from \cref{eq:full_posterior} followed by posterior samples for $Z(\bs)$ one-for-one $\bbeta$, $\btheta$. Next, we estimate gradients and curvatures using the posterior samples of $\phi$, $\sigma^2$ and $\Z$. Posterior sampling is done one-for-one for samples of $\phi$, $\sigma^2$ and $\Z$ using \cref{eq:inference_gradients}. We generated $10,000$ posterior samples, with a $5,000$ burn-in. The model fit was satisfactory: $\widehat{\tau}^2 = 0.384\;(0.145,1.433)$ containing the true value of 1, $\widehat{\sigma}^2 = 344.680\;(194.292, 687.247)$ and $\widehat{\phi}=0.380\;(0.302, 0.489)$, where the highest posterior density (HPD) intervals are shown in the brackets. The fitted process is shown in the left plot of the third row in \Cref{fig:sim-est}, while the estimated gradients and curvatures are shown in the remaining plots of the third and fourth rows of \Cref{fig:sim-est}. We conclude the derivative estimation with \Cref{fig:obs_v_fit} which shows further diagnostics for model fit comparing observed with fitted processes, gradients and curvatures. 

    \begin{figure}[t]
        \centering
        \includegraphics[width=\linewidth]{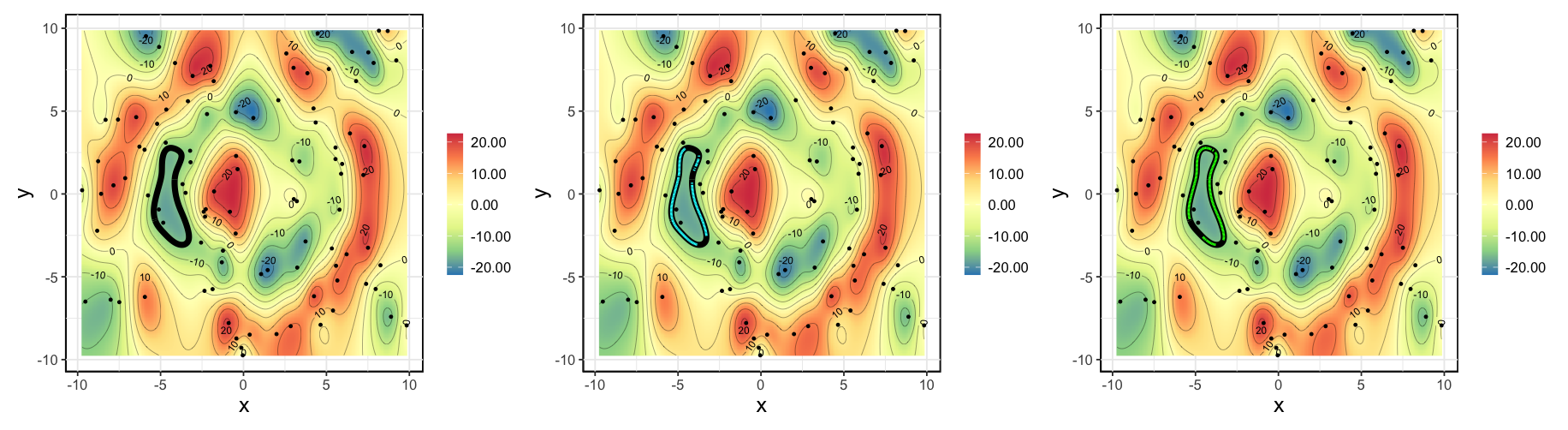}
    \caption{Plots showing (left) the closed curve, $C$ used for wombling (center) line segment level significance for the gradient based wombling measure (right) line segment level significance for the curvature based wombling measure.}
    \label{fig:wombling}
    \end{figure}
    
    The wombling exercise requires a curve. The easiest choice for curves are contours. We choose a closed contour for our purposes. We denote this curve using $C$. Posterior samples of $\bGamma(C)$ are generated one-for-one $\phi$, $\sigma^2$ and $\Z$ using \cref{eq:inference_wms}. The left plot of \Cref{fig:wombling} shows $C$. It is located in a region where the surface is locally increasing. Line segment level inference for gradient and curvature wombling are shown in the top and bottom plots of the right column of \Cref{fig:wombling} respectively. {  For the differential processes the percentage of grid locations that had true values contained in their HPD intervals are as follows: 90\% for $\partial_x$, 92\% for $\partial_y$, 95\% for $\partial^2_{xx}$, 100\% for $\partial^2_{xy}$ and 100\% for $\partial^2_{yy}$.} The higher percentages for the second order derivatives is explained by the increased process variance for the processes $\partial^2_{xx}$, $\partial^2_{xy}$ and $\partial^2_{yy}$ as seen in \Cref{th:2} resulting in wide HPD intervals. Approximately $96\%$ of the line segments had true wombling measures in their corresponding HPD intervals. For the wombling measure, $\widehat{\bGamma(C)}=(-108.765, 154.565)^{\T}$, with 95\% HPD intervals being $(-182.098, -36.290)$ and $(69.053, 241.664)$ respectively, containing the true values $\bGamma(C)=(-131.149, 144.010)^{\T}$. The curve $C$ forms a \emph{wombling boundary} with regard to gradient and curvature. { The significance assessments displayed along the curves are based on marginal posterior credible intervals for individual line segments (not point-wise rather, segment wise, each segment connects two consecutive points in the curve). They should therefore be interpreted as segment-specific posterior assessments rather than as a simultaneous test over the entire curve}. { We compared our approach to \cite{banerjee2006bayesian,halder_bayesian_2024} using software in \cite{Halder_spWombling_Bayesian_Wombling_2023}---under the same settings for posterior inference our approach produced full posterior inference on wombling measures in 2.06 minutes, while the conventional approach took 15.31 minutes. \Cref{fig:compare} shows a comparison between popular methods for clustering using density-based scan statistic (DBSCAN) and $k$-means, we observe that they only offer inference at the observed location and are unable to offer inference over curves. Our chosen curve partially overlaps the cluster (cluster 1 in DBSCAN and region marked with black points in $k$-means).}
    
    { We conclude our experiments by noting that within an ecological setting such curves can either be pre-specified, for example, mountain ranges, rivers or soil and geochemical transitions. Alternatively, they can be chosen by the investigator. Selecting a curve typically requires identifying regions in the spatial domain that sit between high and low abundance, more specifically regions with sharp gradients, which could mark the presence of lurking barriers to the spread of a certain species. We use this principle to select species-specific curves in the next section.}

    \begin{figure}[t]
        \centering
        \includegraphics[width=0.8\linewidth]{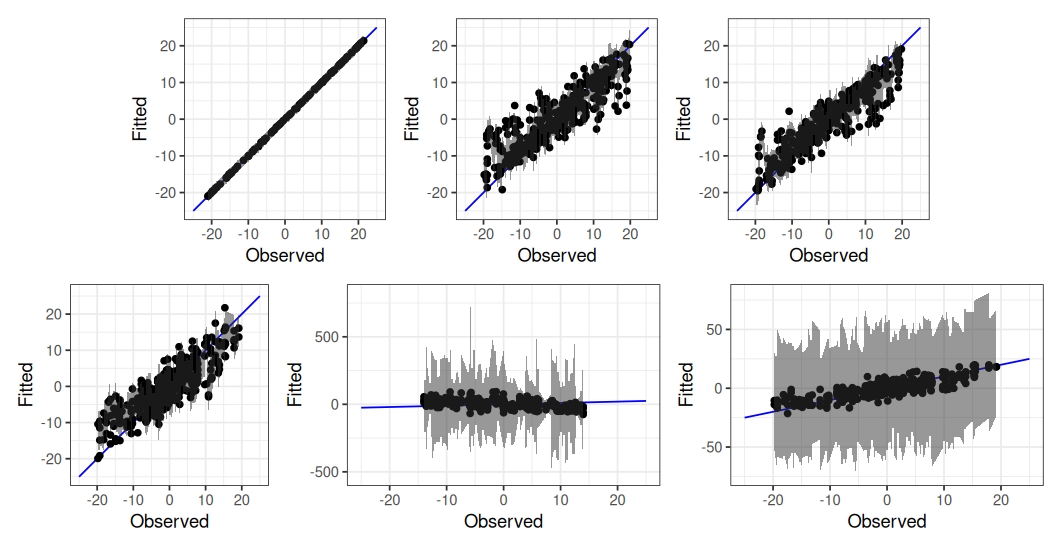}
    \caption{Diagnostics assessing quality of fit using observed vs. fitted values for (top: left) response (top: center) $\partial_x$ (top: right) $\partial_y$ (bottom: left) $\partial^2_{xx}$ (bottom: center) $\partial^2_{xy}$ (bottom: right) $\partial^2_{yy}$ with 95\% credible bands.}
    \label{fig:obs_v_fit}
    \end{figure}

     \begin{figure}[t]
        \centering
        \includegraphics[width=0.95\linewidth]{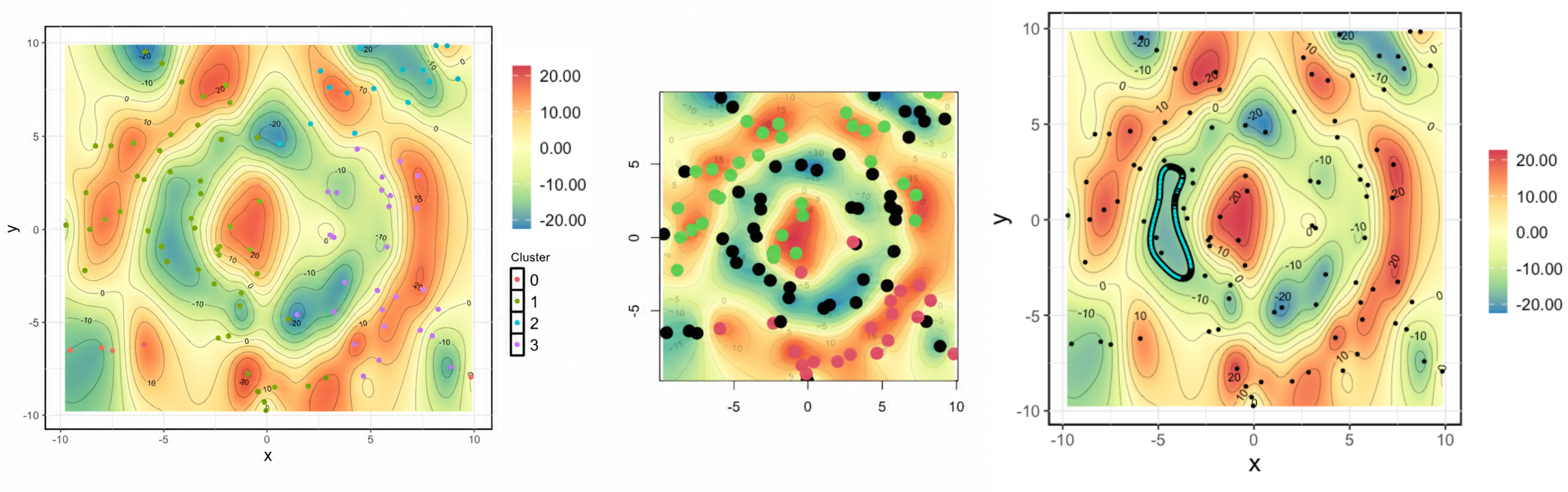}
    \caption{ Plots showing comparison between approaches: (left) hot-spot detection via density based clustering , (center) $k$-means with three mediods (right) the curve chosen for wombling.}
    \label{fig:compare}
    \end{figure}

    \section{Invasive Species in North America}\label{sec:inv_species}

    \begin{table}[t]
\centering
\caption{Parameter and coefficient estimates for different species. The HPD intervals are shown in brackets. Significant effects are marked in bold.}\label{tab:species-bhm}
\resizebox{\linewidth}{!}{
\begin{tabular}{l|@{\extracolsep{10em}}ccc@{}}
  \hline\hline
 $\btheta$, $\bbeta$ &  AT & LD& MV\\ 
  \hline
  \multirow{2}{*}{$\widehat{\phi}$} & 0.789  & 2.408  & 1.835 \\
  & (0.471, 1.190) & (2.048, 2.840) & (1.585, 2.110)\\
  \multirow{2}{*}{$\widehat{\tau}^2$} & 0.583 & 0.571 & 0.508 \\ 
  & (0.516, 0.655) & (0.455, 0.695) & (0.469, 0.555)\\
  \multirow{2}{*}{$\widehat{\sigma}^2$} & 0.307  & 2.454 & 0.551 \\
  & (0.185, 0.496) & (1.796, 3.205) & (0.458, 0.666)\\\hline 
  \multirow{2}{*}{$\widehat{\beta}_0$} & {\bf 0.649} & {\bf 1.972} & {\bf 0.843} \\
  & (0.571, 0.733) & (1.875, 2.064) & (0.795, 0.887)\\
  \multirow{2}{*}{$\widehat{\beta}_{bio7}$} & {\bf 0.009} & {\bf -0.079}  & 0.001 \\ 
  & (0.004, 0.015) & (-0.091, -0.068) & (-0.003, 0.005)\\
  \multirow{2}{*}{$\widehat{\beta}_{bio8}$} & {\bf -0.018}& {\bf 0.132}  & {\bf 0.007} \\ 
  & (-0.028, -0.009) & (0.114, 0.151) & (0.001,  0.013)\\
  \multirow{2}{*}{$\widehat{\beta}_{bio9}$} & -0.006  & -0.016 & {\bf -0.019} \\ 
  & (-0.013, 0.002) & (-0.033, 0.001) & (-0.023, -0.015)\\
   \hline
   \hline
\end{tabular}
}
\end{table}
{
We apply the Bayesian wombling framework to the spatial occurrence of three invasive species of concern: \emph{Adelges tsugae} (AT), \emph{Lycorma delicatula} (LD), and \emph{Microstegium vimineum} (MV). Retaining all three species is useful because they exhibit markedly different degrees and patterns of spatial variation, allowing us to illustrate the behavior of the proposed framework across contrasting ecological settings.

The analysis proceeds in three stages for each species. First, we fit the spatial model in (13) and characterize the residual spatial variation in occurrence after accounting for the bioclimatic covariates. Second, we estimate the local first- and second-order spatial derivative processes. The gradients identify the magnitude and direction of local spatial changes in occurrence, whereas the curvatures characterize changes in those gradients and hence more complex local spatial structure. Third, guided by the estimated spatial surface and its derivatives, we select a candidate curve traversing a region of pronounced spatial change and perform Bayesian wombling. This final step shifts inference from individual locations to the curve itself and evaluates whether the selected curve represents a meaningful spatial boundary.

Because the available data are spatial rather than spatiotemporal, none of these quantities should be interpreted as direct estimates of the temporal movement of an invasion front. Instead, the gradients and wombling boundaries identify sharp spatial transitions in occurrence. When considered together with independent biological and historical information, such transitions may be consistent with invasion boundaries, barriers, corridors, or other features relevant to surveillance and management.

We fit the model
\[
\log\{Y(\mathbf{s})\}
=
\beta_0+\beta_{\mathrm{bio7}}X_{\mathrm{bio7}}
+\beta_{\mathrm{bio8}}X_{\mathrm{bio8}}
+\beta_{\mathrm{bio9}}X_{\mathrm{bio9}}
+Z(\mathbf{s})+\epsilon(\mathbf{s}),
\]
where \(Y(\mathbf{s})\) denotes the species count at location \(\mathbf{s}\), \(Z(\mathbf{s})\) follows a Gaussian process with a Matérn covariance having \(\nu=5/2\), and \(\epsilon(\mathbf{s})\sim N(0,\tau^2)\). The three temperature-related bioclimatic covariates are annual temperature range (bio7), mean temperature of the wettest quarter (bio8), and mean temperature of the driest quarter (bio9).

Table 1 immediately reveals substantial differences among the three species. The proportion of residual variance attributable to the spatial process, \(\sigma^2/(\sigma^2+\tau^2)\), is 34.49\% for AT, 81.12\% for LD, and 52.03\% for MV. Thus, AT exhibits the weakest residual spatial structure, LD the strongest, and MV an intermediate level. These differences provide a useful context for interpreting the derivative and wombling analyses below.

The estimated bioclimatic associations also differ by species. For AT, occurrence is positively associated with bio7 and negatively associated with bio8, whereas for LD these associations are reversed; bio9 is not significant for either species. MV is positively associated with bio8 and negatively associated with bio9, while its association with bio7 is not significant. These species-specific relationships reinforce the need to interpret the subsequent spatial derivative surfaces separately rather than treating the three applications as replications of the same spatial pattern.

\paragraph{Adelges tsugae}

AT provides the case with the weakest residual spatial signal. Consistent with the variance decomposition above, the estimated spatial effects in Figure 6 are comparatively localized, with significant spatial structure concentrated primarily around the New York–New Jersey–Connecticut region, the Carolinas, and Nova Scotia. The derivative surfaces similarly contain relatively few significant locations. Significant longitudinal and latitudinal gradients occur primarily along portions of the eastern United States, the Carolinas, and Nova Scotia, indicating localized rather than widespread sharp spatial transitions in occurrence. Significant curvature is still more restricted, although the mixed-curvature surface suggests comparatively complex local spatial structure around New York.

For the curve-level analysis, we select a curve traversing the tri-state region, shown in the first row of Figure 9. This region was chosen because the fitted derivative surfaces indicate localized spatial change there. The gradient wombling analysis identifies significant segments primarily along the northern portion of the curve near Connecticut, whereas there is substantially less evidence of significant directional curvature. The estimated total wombling measures are
\[
\widehat{\Gamma}(C)=(0.219,-0.188)^T,
\]
with 95\% HPD intervals \((-0.002,0.465)\) for the gradient measure and \((-0.862,0.506)\) for the curvature measure. Thus, the curve-level evidence for AT is relatively weak compared with the other species: the gradient measure is close to significance, while the curvature measure clearly includes zero. This result is consistent with the comparatively weak residual spatial structure estimated for AT and illustrates a setting in which wombling identifies localized candidate boundary segments without strong evidence for a boundary over the entire selected curve.

\paragraph{Lycorma delicatula}

LD presents a sharply contrasting case. It has the largest spatial variance component among the three species, with approximately 81\% of residual variation attributable to the spatial process. Correspondingly, Figure 7 shows pronounced spatial structure concentrated around the tri-state region, Pennsylvania, Delaware, and adjoining portions of the eastern United States.

The estimated derivative surfaces reveal stronger rates of spatial change than those observed for AT. Significant longitudinal gradients occur in Ohio, Pennsylvania, and Virginia, indicating pronounced west–east transitions in occurrence, while significant latitudinal gradients are apparent along portions of the coastal region. The magnitudes of the estimated gradients and curvatures are also larger than for the other two species. Although these gradients describe spatial rather than temporal change, their orientation is consistent with independent studies reporting predominantly westward spread and an over-representation of westward dispersal jumps for LD. The agreement therefore provides useful external context for the spatial transitions identified by the model without requiring that the gradients themselves be interpreted as estimates of temporal spread.

We select a curve traversing New York, New Jersey, and Pennsylvania for the wombling analysis (Figure 9, second row). In contrast to AT, a large proportion of the curve consists of segments with significant directional gradients, providing substantially stronger evidence of a spatial boundary. The total wombling measure is
\[
\widehat{\Gamma}(C)=(6.398,-6.677)^T,
\]
with 95\% HPD intervals \((4.346,8.705)\) and \((-20.543,7.758)\) for the gradient and curvature measures, respectively. The gradient interval excludes zero, establishing the selected curve as a wombling boundary with respect to first-order spatial change. The curvature interval, however, includes zero. Thus, the principal evidence for the LD boundary comes from a persistent directional gradient across the curve rather than from an integrated second-order change.

\paragraph{Microstegium vimineum}

MV represents an intermediate case in terms of the strength of residual spatial dependence but differs from both insect species in its geographic extent. Significant spatial effects are distributed over a broader portion of the eastern and southern United States (Figure 8), consistent with its wider observed geographic distribution.

The derivative surfaces show significant gradients and curvatures along portions of the eastern coast and southern United States. In particular, the first-order surfaces identify pronounced westward and southward spatial changes in occurrence. Unlike AT, where significant derivative behavior is comparatively localized, the MV results indicate spatial transitions distributed across a broader geographic region.

For wombling, we select an open curve along the eastern region shown in the third row of Figure 9. Most segments along this curve exhibit significant directional changes. The total wombling measure is
\[
\widehat{\Gamma}(C)=(3.013,-6.891)^T,
\]
with 95\% HPD intervals \((1.982,4.133)\) for the gradient measure and \((-11.997,-1.953)\) for the curvature measure. Both intervals exclude zero. Consequently, the selected curve constitutes a wombling boundary with respect to both first- and second-order spatial change. This distinguishes MV from LD, for which the integrated evidence is primarily gradient-based, and from AT, for which the curve-level evidence is considerably weaker.

Taken together, the three applications demonstrate three distinct inferential regimes. AT exhibits comparatively weak spatial dependence and only localized evidence of sharp spatial change; LD exhibits strong spatial dependence and a pronounced gradient-based boundary; and MV exhibits intermediate overall spatial dependence but a geographically broader pattern of spatial change, with evidence for both gradient- and curvature-based wombling. The three species therefore provide complementary rather than redundant illustrations of the proposed framework.
}

    \begin{figure}[t]
    \centering
    \includegraphics[width=0.95\linewidth]{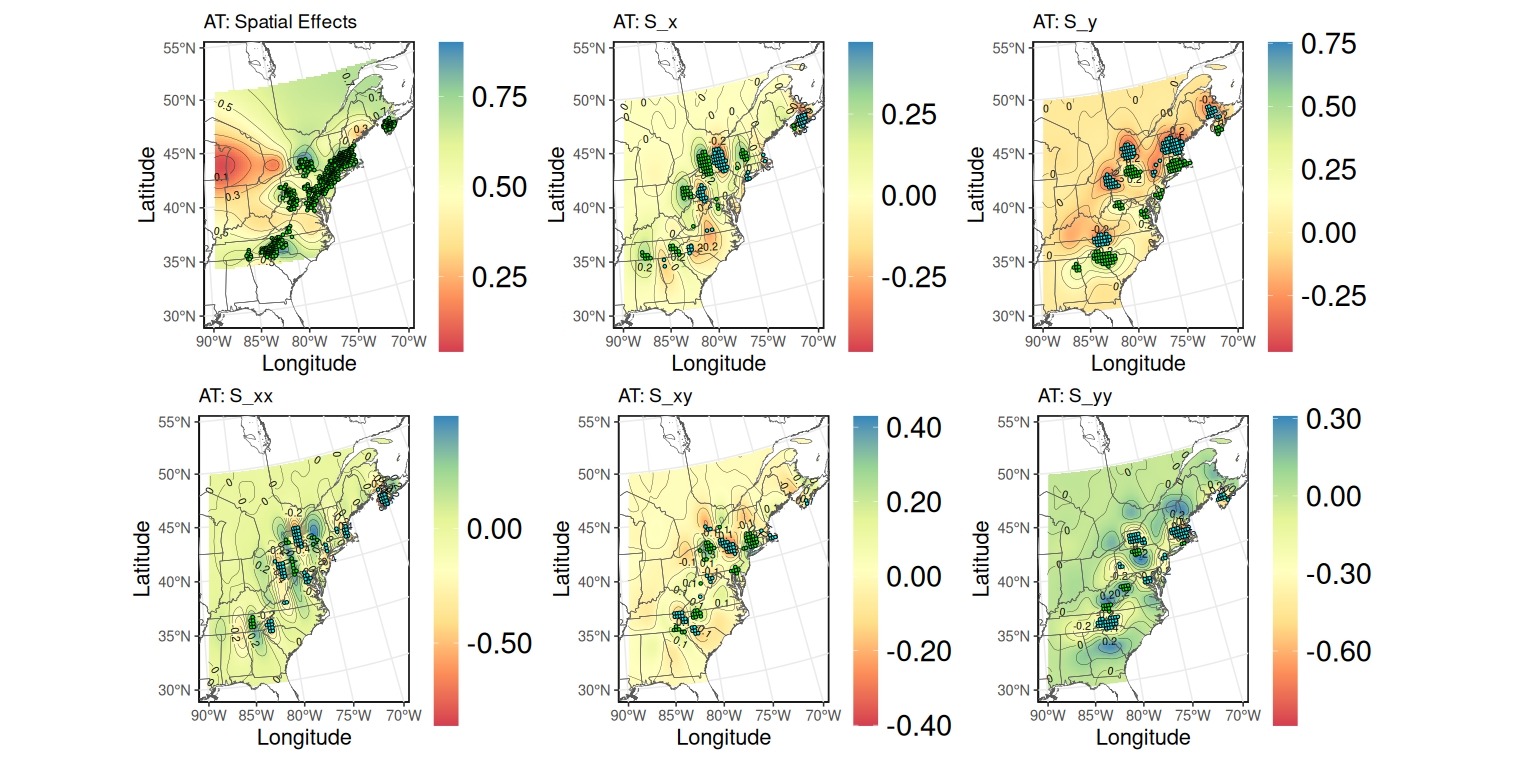}
    \caption{ Interpolated spatial plots for AT showing (row 1) estimated spatial effect, estimated gradient $\partial_x$ along the longitude, estimated gradient $\partial_y$ along the latitude; (row 2) estimated curvature, $\partial^2_{xx}$ along the longitude, estimated mixed curvature, $\partial^2_{xy}$ and estimated curvature, $\partial^2_{yy}$ along the latitude. Only significant locations and grid points are shown--positive with green and negative with cyan.}
    \label{fig:species-grads-at}
    \end{figure}

    \begin{figure}[t]
    \centering
    \includegraphics[width=0.95\linewidth]{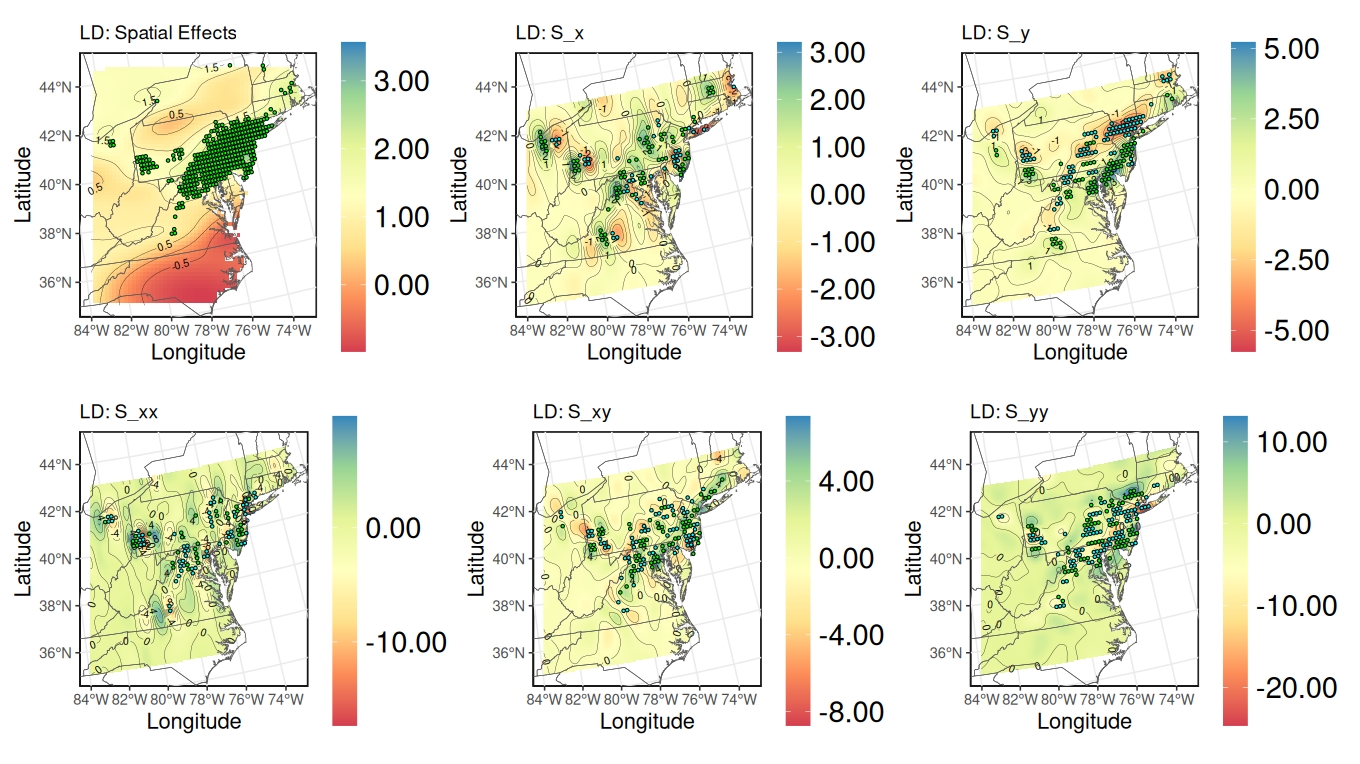}
    \caption{Interpolated spatial plots for LD showing (row 1) estimated spatial effect, estimated gradient $\partial_x$ along the longitude, estimated gradient $\partial_y$ along the latitude; (row 2) estimated curvature, $\partial^2_{xx}$ along the longitude, estimated mixed curvature, $\partial^2_{xy}$ and estimated curvature, $\partial^2_{yy}$ along the latitude. Only significant locations and grid points are shown--positive with green and negative with cyan.}
    \label{fig:species-grads-ld}
    \end{figure}

    \begin{figure}[t]
    \centering
    \includegraphics[width=0.95\linewidth]{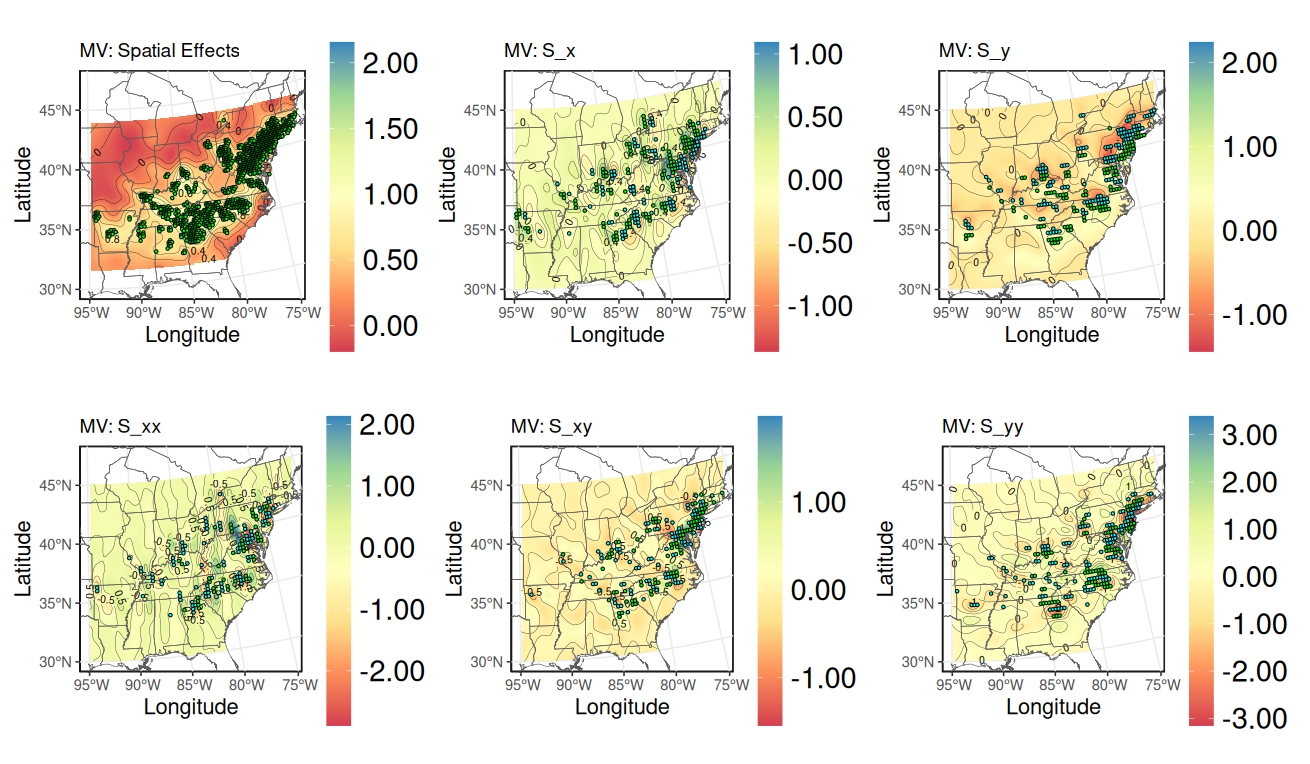}
    \caption{Interpolated spatial plots for MV showing (row 1) estimated spatial effect, estimated gradient $\partial_x$ along the longitude, estimated gradient $\partial_y$ along the latitude; (row 2) estimated curvature, $\partial^2_{xx}$ along the longitude, estimated mixed curvature, $\partial^2_{xy}$ and estimated curvature, $\partial^2_{yy}$ along the latitude. Only significant locations and grid points are shown--positive with green and negative with cyan.}
    \label{fig:species-grads-mv}
    \end{figure}

    \begin{figure}[t]
        \centering
        \includegraphics[width=\linewidth]{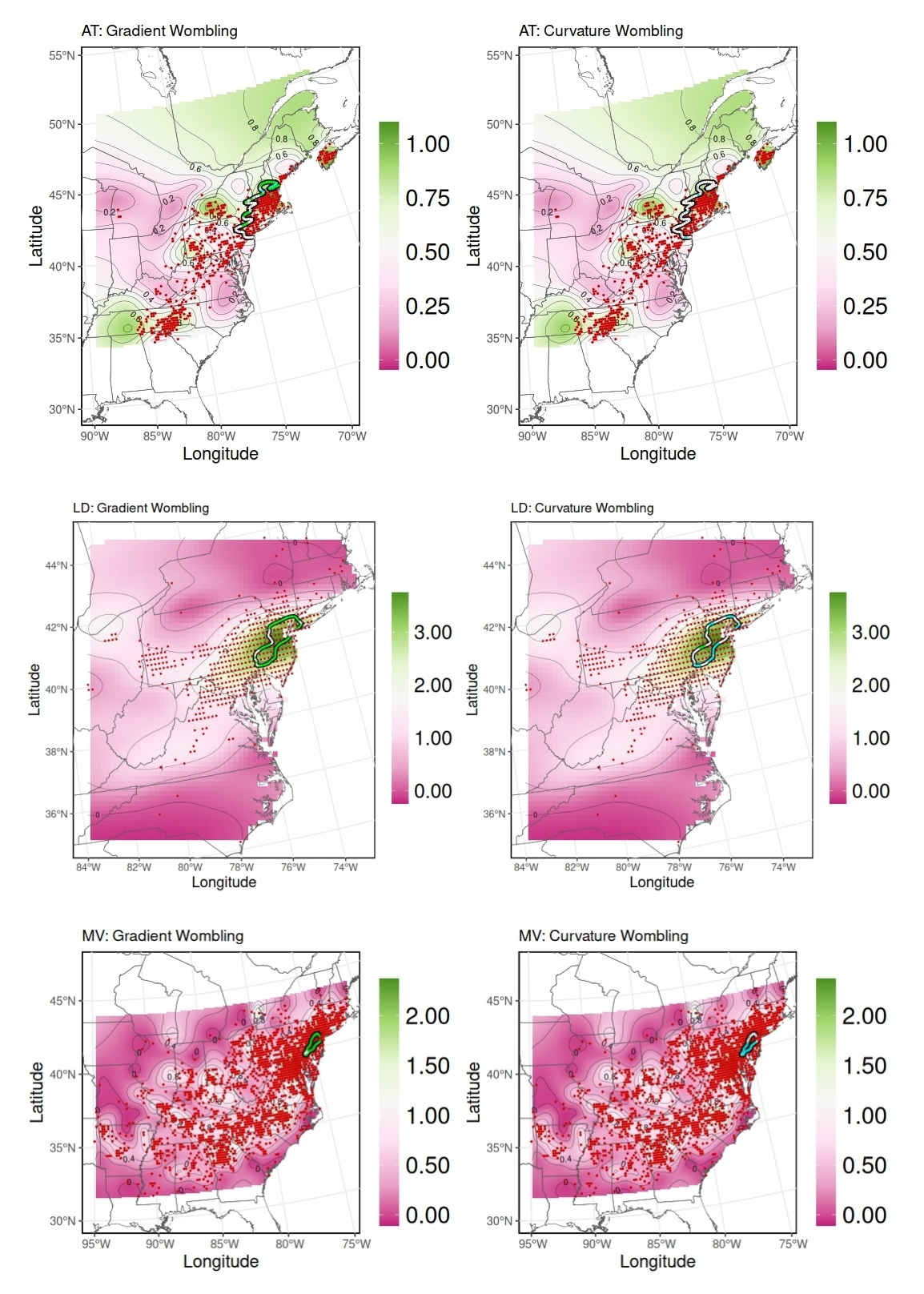}
        \caption{Spatial plots showing line-segment level inference for gradient (left column) and curvature (right column) wombling respectively for the three species. Significant segments of curves are highlighted--positive with green and negative with cyan.}
        \label{fig:species-wm}
    \end{figure}

    \section{Discussion and Future Work}\label{sec:diss}
    { We have demonstrated the use of Bayesian wombling for detecting regions of rapid spatial change in the occurrence of invasive species. Our framework provides Bayesian inference over curvilinear boundaries and quantifies spatial gradients and curvatures in species occurrence within a hierarchical model that also incorporates bio-climatic covariates. Importantly, because the data analyzed here are spatial rather than spatiotemporal, these gradients characterize spatial variation in occurrence and should not be interpreted as direct estimates of temporal spread or movement of an invasion front. Instead, the detected boundaries identify sharp spatial transitions that may represent invasion boundaries, barriers to spread, corridors, environmental transitions, or other features of the realized species distribution. For some species, the orientation of the identified gradients is consistent with independently observed and modeled patterns of spread, providing biological context for the spatial patterns detected by the model.}
    
    There are many avenues for future research. { Further enhancements can be made to testing the curve for a wombling boundary via multiple testing. Multiple candidate boundaries can be considered and $\Gamma(C)$ can be maximized over such boundaries to yield a final choice.} We assume that the logarithm of counts is a Gaussian Process which is not ideal considering the data is originally a count. Similar frameworks can be developed for generalized linear models, particularly focusing on zero-inflated models \citep[see, e.g.][]{halder_spatial_2020,halder2021spatial,halder_bayesian_2023} which provide a more realistic setting for analyzing raw presence/absence counts. Developing Bayesian wombling using generalized linear models (GLMs) would be beneficial in that regards. Extending the methodology to forecast and back-cast the invasion front would support risk management decisions (e.g. decision to eradicate, contain, or monitor a pest), as well as supporting forensic efforts to identify the initial point of establishment for an invasive species. { Direct inference on the movement of an invasion front through time would require spatiotemporal occurrence data and lies beyond the spatial framework considered here.} This would require extending the spatiotemporal framework in \cite{halder_bayesian_2026} to GLMs. Additionally, applying this methodology to past invasions could identify whether pest management actions were effective in managing the pest, especially in the case of 'slow-the-spread' efforts, and provide a decision support tool in managing future pest invasions. Another promising approach is to include altitude in the spread model and perform boundary analysis over a curved geometry \citep[see, e.g.][]{li_statistical_2026}. While our approach employs MCMC for model fitting, we recognize opportunities to incorporate alternative, possibly faster, machine learning tools such as Bayesian predictive stacking \citep[see, for e.g.,][]{pan_bayesian_2025,zhang_bayesian_2026} that are implemented in the \emph{spStack} \citep[see, for e.g.,][]{pan2024spstack} for wombling using stacked posterior surfaces. Extending these models to a multivariate setting will allow us to track correlated regions of spread across multiple species. Finally, an open-source software for ecologists can follow the steps laid out in \cite{halder_nimblewomble_2025,halder_rj_2026}.

    \section{Acknowledgements}
    AH acknowledges funding from the National Science Foundation, Division of Mathematical Sciences (DMS 2515897), the National Institutes of Health, National Institute of Environmental Health Sciences (R56 ES036250, R21 CA305033), and institutional funds at the Dornsife School of Public Health and the Urban Health Collaborative at Drexel University.

\bibliographystyle{apalike}
\bibliography{main}

\end{document}